\documentclass[reprint, superscriptaddress, amsmath, amssymb, prl, notitlepage, longbibliography, floatfix, twocolumn]{revtex4-1}

\usepackage{tensor}     
\usepackage{graphicx}   
\usepackage{subfigure} 
\usepackage[
colorlinks=true,        
citecolor=blue,         
linkcolor=blue,         
urlcolor=blue           
]{hyperref}             
\usepackage{bm}         
\usepackage{xcolor}     

\newcommand{\newc}{\newcommand*} 
\newc{\figurewidth}{3.2in}
\newc{\xbar}{\bar{x}}
\newc{\rhoeq}{\rho_{\rm{eq}}}
\newc{\zeq}{z_{\rm{eq}}}
\newc{\la}{\lambda}
\newc{\tla}{\tilde{\la}}
\newc{\dt}{\delta}
\newc{\Dt}{\Delta}
\newc{\vj}{\vec{j}}
\newc{\vl}{\vec{l}}
\newc{\hx}{\hat{x}}
\newc{\hy}{\hat{y}}
\newc{\bj}{\bm{j}}
\newc{\mJ}{\mathcal{J}}
\newc{\mP}{\mathcal{P}}
\newc{\ga}{\gamma}
\newc{\Msun}{M_\odot}
\newc{\app}{\approx}
\newc{\av}[1]{\langle #1 \rangle}
\newc{\eq}[1]{Eq.~\eqref{#1}}
\newc{\al}{\alpha}
\newc{\Xstar}{X_{\ast}}
\newc{\seq}{\sigma_{\rm{eq}}}
\newc{\fpbh}{f_{\rm{pbh}}}
\newc{\VT}{\langle VT \rangle}

\def\({\left(}
\def\){\right)}
\def\[{\left[}
\def\]{\right]}

\def\e{\begin{equation}}
\def\q{\end{equation}}
\def\m{\begin{eqnarray}}
\def\n{\end{eqnarray}}

\begin{document}

\title{Oscillons during Dirac-Born-Infeld Preheating}

\author{Yu Sang}
\email{sangyu@yzu.edu.cn}
\affiliation{Center for Gravitation and Cosmology, College of Physical Science and Technology, Yangzhou University, Yangzhou 225009, China}
\affiliation{Shanghai Frontier Science Research Center for Gravitational Wave Detection, School of Aeronautics and Astronautics, Shanghai Jiao Tong University, Shanghai 200240, China}
\author{Qing-Guo Huang}
\email{Corresponding author: huangqg@itp.ac.cn}
\affiliation{CAS Key Laboratory of Theoretical Physics, 
Institute of Theoretical Physics, Chinese Academy of Sciences,
Beijing 100190, China}
\affiliation{School of Physical Sciences, 
University of Chinese Academy of Sciences, 
No. 19A Yuquan Road, Beijing 100049, China}
\affiliation{School of Fundamental Physics and Mathematical Sciences
Hangzhou Institute for Advanced Study, UCAS, Hangzhou 310024, China}

\date{\today}
\begin{abstract}
Oscillons are long-lived, localized, oscillating nonlinear excitations of a real scalar field which can be abundantly produced during preheating after inflation. We give the first $(3+1)$-dimensional simulation for the oscillon formation during preheating with noncanonical kinetic terms, e.g. the Dirac-Born-Infeld form, and find that the formation of oscillons is significantly influenced by the noncanonical effect.  
\end{abstract}

\pacs{???}

\maketitle

{\it Introduction.} An interesting possible observable effect from preheating is the long-lived, localized, oscillating nonlinear excitation of a real scalar field, so-called oscillon \cite{Bogolyubsky:1976yu,Gleiser:1993pt,Copeland:1995fq,Copeland:2002ku,Gleiser:2006te,Hindmarsh:2006ur,Saffin:2006yk,Graham:2006vy,Gleiser:2008ty,Amin:2010jq,Amin:2010dc,Amin:2011hj,Antusch:2017flz,Antusch:2019qrr,Kou:2019bbc,Nazari:2020fmk}. During preheating after inflation, oscillons are abundantly produced  following efficient self-resonance of inflaton field if the potential is quadratic at the bottom and flattening away from the minimum \cite{Amin:2010jq}. The formation of oscillons in the postinflationary Universe has significant cosmological consequences, e.g., the existence of an oscillon-dominated phase \cite{Gleiser:2010qt,Gleiser:2011xj,Amin:2011hj,Lozanov:2014zfa,Lozanov:2016hid,Lozanov:2017hjm,Lozanov:2019ylm} and the production of gravitational wave background \cite{Zhou:2013tsa,Antusch:2016con,Antusch:2017vga,Liu:2017hua,Liu:2018rrt,Amin:2018xfe,Amin:2018xfe,Lozanov:2019ylm,Sang:2019ndv,Hiramatsu:2020obh}.

Although most of the literature concentrates on oscillons of a canonical scalar field, it is still interesting to probe oscillons in the scenario of noncanonical scalar field. In particular, brane inflationary models based on string theory have been extensively studied in the literature \cite{Dvali:1998pa,Dvali:2001fw,Kachru:2003sx,Burgess:2001fx,Shiu:2001sy}. The kinetic term for the inflaton field associated with the motion of a D3-brane in higher-dimensional background spacetime naturally takes the Dirac-Born-Infeld (DBI) form which reduces to the canonical kinetic term in the low velocity limit. During preheating after brane inflation, the inflaton field moves fastly and it is natural to take the full DBI form into account, and the phenomenological consequences of the nonlinear dynamics during preheating after brane inflation provide a possible way to explore string theory.

 Preheating with noncanonical kinetic terms has been studied in literature. See some related works in \cite{Lachapelle:2008sy,Matsuda:2008hk,Bouatta:2010bp,Karouby:2011xs,Child:2013ria,Amin:2013ika}. In particular, lattice simulation on preheating with parametric resonance has been extended to the case of a DBI inflaton field coupled to a canonical matter field in \cite{Child:2013ria}, and the authors showed that the parametric resonance in the matter field and self-resonance in the inflaton field are as efficient as in traditional preheating. In \cite{Amin:2013ika}, the existence of oscillons in scalar field theories with noncanonical kinetic terms  in the small-amplitude limit can be supported solely by the noncanonical kinetic terms, without any need for nonlinear terms in the potential.

 In this letter, we will give the first fully $(3+1)$-dimensional lattice simulation for the formation of oscillons during DBI preheating, and our numerical results indicate that the noncanonical effect can significantly influence the formation of oscillons.

{\it Model and method.} Let's start with a general form of Lagrangian for the noncanonical inflaton field, 
\begin{equation}\label{eq:GenLag}
S=\frac{1}{2}\int d^4x \sqrt{-g} \left[M_{pl}^2R +2P(X,\phi)\right]~,
\end{equation}
where $\phi$ is the inflaton field, $X=-\frac{1}{2}g^{\mu\nu}\partial_{\mu}\phi \partial_{\nu}\phi$, and $M_\mathrm{pl}=1/\sqrt{8\pi G}$ is the reduced Planck mass. 
In a Friedmann-Lema\^itre-Robertson-Walker (FLRW) universe
\begin{equation}\label{eq:FRW}
ds^2=-dt^2+a^2(t) \delta_{ij}  dx^i dx^j~,
\end{equation}
the equation of motion of $\phi$ is 
\begin{eqnarray}\label{eq:phi}
\nonumber
&&P_{,X} \ddot{\phi}+ 3\frac{\dot{a}}{a}P_{,X} \dot{\phi} + \partial_0 (P_{,X})\dot{\phi} - \frac{1}{a^2} P_{,X} \nabla^2\phi   \\ 
 && - \frac{1}{a^2} \partial_i (P_{,X}) \partial_i\phi -P_{,\phi}=0 ~, 
\end{eqnarray}
where $P_{,X}$ denotes the derivative with respect to $X$ and $P_{,\phi}$ denotes the derivative with respect to $\phi$.

In this letter, based on string theory, we consider the DBI inflation \cite{Silverstein:2003hf,Alishahiha:2004eh} with 
\begin{equation}\label{eq:P-DBI} 
P(X,\phi) = -\frac{1}{f(\phi)}\left(\sqrt{1-2Xf(\phi)} -1\right)- V(\phi) ~,
\end{equation}
which is motivated by brane inflationary models in warped compactifications. The Lagrangian is a low energy description of a D3 brane travelling in a warped throat. The inflaton field $\phi$ is a collective coordinate which measures the radial position of the brane in the throat. The potential $V(\phi)$ describes how the brane is attracted towards the bottom of the throat. The warp factor $f(\phi)$ is determined by the warped geometry. For an anti-de Sitter throat,  $f(\phi)$ takes the form 
\begin{equation}\label{eq:warp}
f(\phi)=\frac{\lambda}{(\phi^2+\mu^2)^2}~,
\end{equation}
where $\mu$ is a parameter corresponding to the infrared cutoff scale. 
The equation of motion of $\phi$ is given by 
\begin{eqnarray}\label{eq:phi-DBI}
\nonumber
&&\ddot{\phi}\left( 1+\frac{f}{a^2} (\partial_i \phi)^2 \right) + 3\frac{\dot{a}}{a}\frac{\dot{\phi}}{\gamma^2} - \frac{\nabla^2 \phi}{a^2} 
-\frac{f_{,\phi}}{f^2}+\frac{3f_{,\phi} X}{f} \\ 
\nonumber
 &&+ \frac{1}{\gamma^3}\left( \frac{f_{,\phi}}{f^2} + V_{,\phi} \right)  + \frac{f}{a^2}  \Bigg[  \left(\frac{\dot{a}}{a}\dot{\phi} - \frac{\nabla^2\phi}{a^2} \right)  (\partial_i \phi)^2 \\  
 &&+ \nabla^2\phi \dot{\phi}^2 + \frac{\partial_i\phi\partial_j \phi \partial_j \partial_i \phi}{a^2} - 2 \dot{\phi} \partial_i \phi  \partial_i \dot{\phi} \Bigg]=0~, 
\end{eqnarray}
where the Lorentz factor $\gamma \equiv \left(1-2Xf\right)^{-1/2}$, $f_{,\phi} \equiv \frac{df}{d\phi}$ and $V_{,\phi} \equiv \frac{\partial V}{\partial\phi}$.
In this letter, we consider the potential of inflaton $\phi$ as follows 
\begin{equation}\label{eq:axionV}
V(\phi)=m^2M^2 \[\(1+{\phi^2\over M^2}\)^{1/2}-1\]~,
\end{equation}
which is motivated by the axion monodromy inflation \cite{McAllister:2008hb,Silverstein:2008sg}. Self-resonance of a canonical inflaton field with such a potential leads to copious oscillon generation \cite{Amin:2011hj} and gravitational waves are significantly produced when oscillons are being formed \cite{Zhou:2013tsa,Sang:2019ndv}.

The expansion of the Universe is driven by the energy density of the inflaton field.
The evolution of scale factor is described by the Friedmann equation
\begin{equation}\label{eq:Fried}
 \left( \frac{\dot{a}}{a} \right)^2 =\frac{1}{3M_\mathrm{pl}^2 } \rho ~,
\end{equation}
where $\rho$ is the energy density averaged over the spatial volume
\begin{equation}\label{eq:rho}
\rho \equiv \langle T_{00} \rangle = \left\langle  \frac{1}{f} (\gamma^{-1}-1) + \gamma \dot{\phi}^2 + V\right\rangle~.
\end{equation}

We use a (3+1)-dimensional numerical simulation to study the preheating and oscillon formation during DBI preheating after inflation.
Our code is based on a publicly available lattice evolution program, Grid and Bubble Evolver (GABE) \cite{GABEwebpage,Child:2013ria}, which has been well-used in the preheating scenarios \cite{Deskins:2013dwa,Adshead:2015pva,Adshead:2017xll,Nguyen:2019kbm,Giblin:2019nuv}. Using a second-order Runge-Kutta method for time integration, GABE stores the field and field derivative value at the same times during the time step, which is required by terms such as the product of the field with its time derivative. For spatial derivative terms, GABE uses the second-order finite differencing, and we further update the code to fourth-order centered differencing scheme. We use a $N=256^3$ box in the lattice simulations. The parameters in potential are set to $m=3.1 \times 10^{-5} M_{\rm{pl}}$ and $M=0.01M_{\rm{pl}}$, and parameter in the warp factor $f$ is $\mu = 5\times 10^3 M_{\rm{pl}}$. We consider two cases of $\lambda = 3\times10^{26}$ and $1.0 \times 10^{27}$ respectively. We will show that the space-averaged value of Lorentz factor $\gamma$ has a maximum value of  2.075 for $\lambda = 3\times10^{26}$ and 3.898 for $\lambda = 1.0 \times 10^{27}$ in the simulation, which means the inflaton field significantly departs from canonical case. 

In principle, the scalar field could be the DBI form during both inflationary and preheating phases. In order to show up the noncanonical effects on the oscillons only during preheating phase, not the inflationary phase, by comparing our results with those in canonical form in the whole dynamics, we prepare the scalar configuration in the inflationary phase to be slowly rolling. Therefore, the preheating follows a slow-roll inflation with $\gamma = 1$ but the noncanonical terms begin to be important after inflation. We evolve the homogeneous inflaton field from slow-roll region, which gives a constraint on the field and its derivative. Using this constraint as an initial condition, we numerically solve the equation of motion of the homogeneous inflaton field, i.e., the Eq. (\ref{eq:phi-DBI}) without the gradient terms, to determine the initial values of inflaton field and its derivative for our lattice simulation. The initial values are $\phi_i = 0.50 M_{\rm{pl}}$ and $\dot{\phi}_i \approx -1.17\times 10^{-6}M^2_{\rm{pl}}$ for the case of $\lambda = 3\times10^{26}$ and $\phi_i = 0.50 M_{\rm{pl}}$ and $\dot{\phi}_i \approx -7.42\times 10^{-7}M^2_{\rm{pl}}$ for the case of $\lambda = 1.0 \times 10^{27}$. As a contrast, we also consider the canonical case in our simulation,  initial values of which are determined as $\phi_i = 0.50 M_{\rm{pl}}$ and $\dot{\phi}_i \approx -1.86\times 10^{-6}M^2_{\rm{pl}}$. Those initial values of the  inflaton field and its derivatives ensure the same initial horizons in lattice simulations for all the models. As an approximation, the fluctuations of field and its derivative are initialized as the Bunch-Davies vacuum (see Ref. \cite{Child:2013ria} for details). Since preheating is insensitive to the initial power spectrum \cite{Frolov:2008hy}, we expected the initial conditions of the fluctuations will not affect the results of the simulation.

\begin{figure}[!htbp]
\centering
\includegraphics[width=0.45\textwidth]{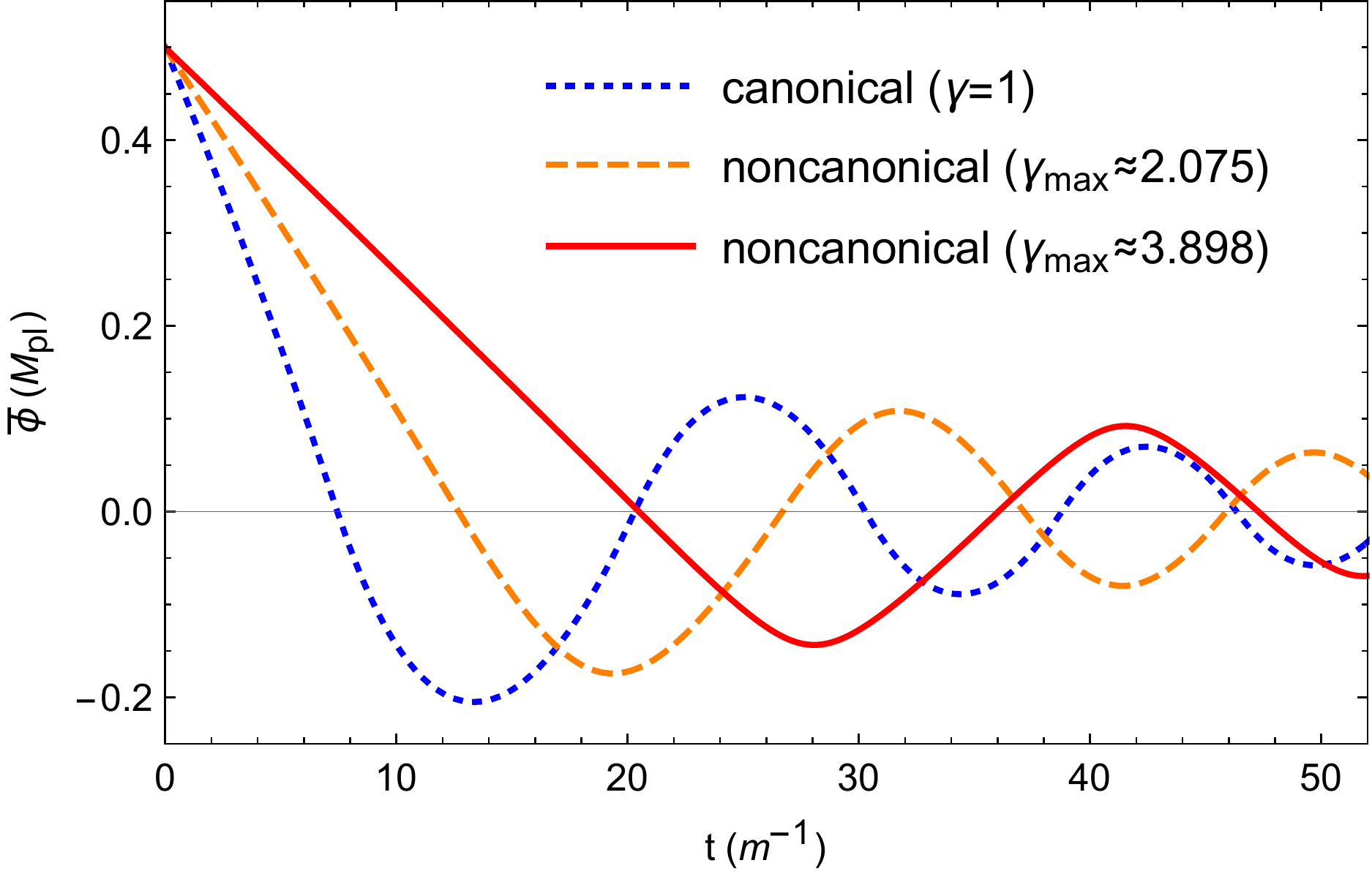}
\caption{Evolution of the mean inflaton field for simulations with $\mu = 5\times 10^3 M_{\rm{pl}}$, $\lambda=1.0 \times 10^{27}$, $\gamma_{\rm{max}} \approx 3.898$ (red solid line), with $\mu = 5\times 10^3 M_{\rm{pl}}$, $\lambda=3\times10^{26}$, $\gamma_{\rm{max}} \approx 2.075$ (orange dashed line) and canonical case $\gamma = 1$ (blue dotted line). }
\label{fig:phi_evl} 
\end{figure}

{\it Results. } The noncanonical kinetic terms bring terms with $\gamma$ factor and higher order derivatives into the equation of motion of inflaton field. The effect of these terms on the evolution of the field can be  illustrated by the mean inflaton field value shown in Fig.~\ref{fig:phi_evl}. 
From Fig.~\ref{fig:phi_evl}, we see that the noncanonical kinetic term can significantly change the evolution of the inflaton field during DBI preheating. As the inflaton field and its velocity oscillate with time, the relativistic factor $\gamma$ also exhibits oscillatory behavior. 
The space-averaged value of $\gamma$ arrives its maximum value $\gamma_{\rm{max}} \approx 2.075$ at  $t \approx 12.4 m^{-1} $ when field has largest velocity in our setup for the case of $\lambda = 3\times10^{26}$, and $\gamma_{\rm{max}} \approx 3.898$ at  $t \approx 20 m^{-1} $ for the case of $\lambda = 1.0 \times 10^{27}$.  Due to the presence of noncanonical terms, the motion of inflaton field approaches a sawtooth pattern, unlike the canonical field with a sinusoidal evolution. As the noncanonical terms become efficient, the sawtooth grows sharper, which enhance the resonance. However, the lengthened period of oscillations suppresses resonance and these two effects compete with each other. According to Ref. \cite{Karouby:2011xs}, the second effect should dominate over the former one, leading to a less efficient preheating for the noncanonical inflaton field. The result of self-resonance here is consistent with the simulation in  Ref. \cite{Child:2013ria}.

\begin{figure}[!htbp]
\centering
\includegraphics[width=0.45\textwidth]{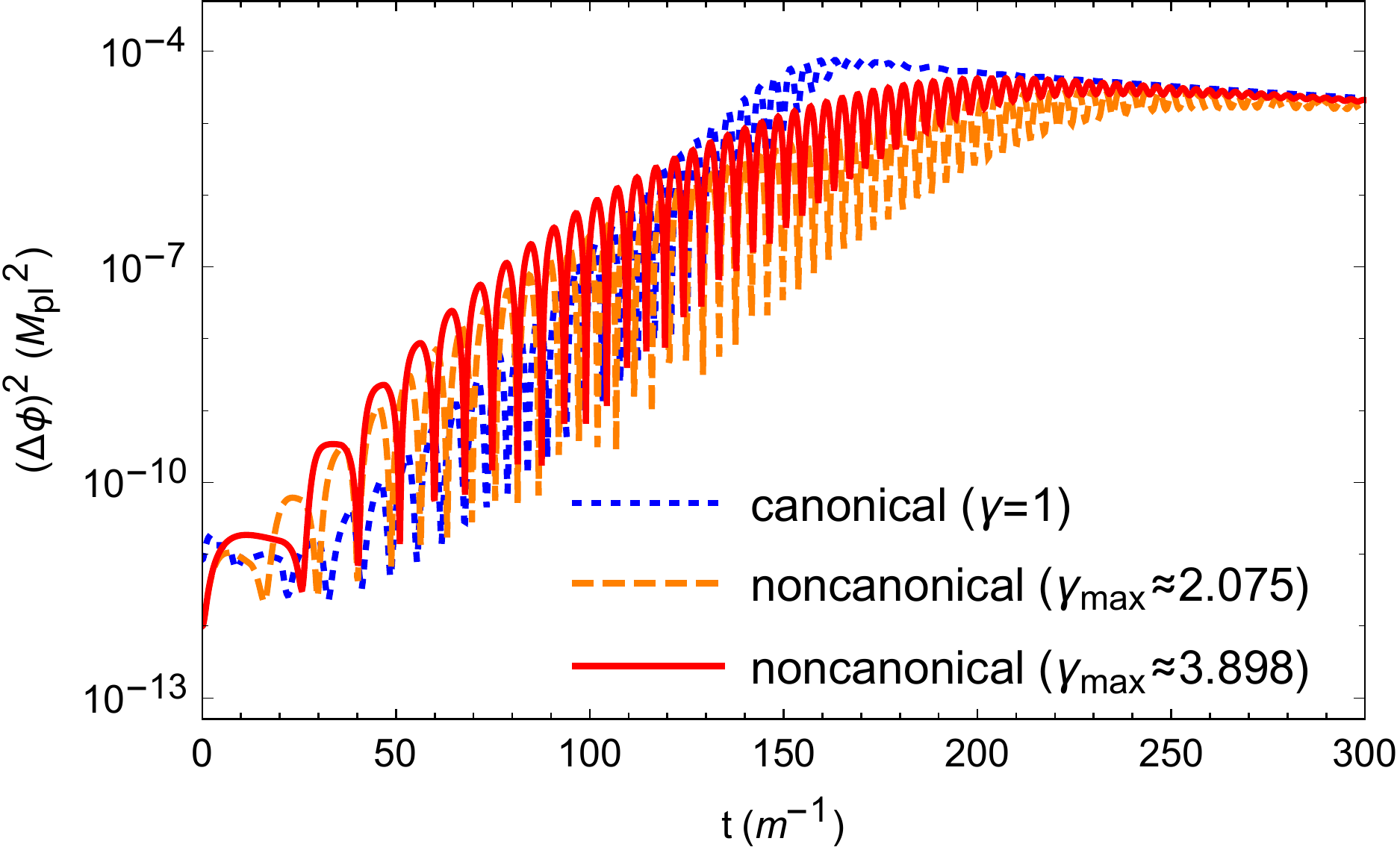}
\caption{Evolution of the variance of the inflaton field for simulations with $\mu = 5\times 10^3 M_{\rm{pl}}$, $\lambda=1.0 \times 10^{27}$, $\gamma_{\rm{max}} \approx 3.898$ (red solid line), with $\mu = 5\times 10^3 M_{\rm{pl}}$, $\lambda=3\times10^{26}$, $\gamma_{\rm{max}} \approx 2.075$ (orange dashed line) and canonical case $\gamma = 1$ (blue dotted line). }
\label{fig:phi_var} 
\end{figure}

The evolution of variance of the inflaton field is shown in Fig.~\ref{fig:phi_var}. 
By comparing the time of occurrence, we see the sharp troughs in the evolution in Fig.~\ref{fig:phi_var} correspond to the extreme points of motion of inflaton field in Fig. \ref{fig:phi_evl}, and the peaks to the zero points, for both noncanonical and canonical cases, except for the initial part of the first oscillation period ($t \lesssim 20 m^{-1} $). This implies that inflaton field with larger velocity tends to be more inhomogeneous in a local temporal region. For example, in the case of $\lambda = 1.0 \times 10^{27}$, during the evolution of noncanonical inflaton field  from  $t \approx 41 m^{-1} $ to $t \approx 47 m^{-1} $, the velocity of inflaton field varies from zero to its local maximum. Meanwhile, the variance of field grows from $6.8\times 10^{-12}M_{\rm{pl}}^2$ to $2.3\times 10^{-9}M_{\rm{pl}}^2$. Unlike the case of inflaton field coupled with a scaler matter field, in which self-resonance of noncanonical and canonical inflaton field have dramatic difference (see Fig. 3 in Ref. \cite{Child:2013ria}),  here for the single inflaton preheating with potential (\ref{eq:axionV}), we see the variance of $\phi$ grows quickly in the early stage for both noncanonical and canonical cases. The reason is that the mechanism for self-resonance is provided by the potential term, and for the case with noncanonical terms, they suppress the efficiency of self-resonance. For example, the noncanonical terms delay the end of self-resonance until approximately  $t \sim 200 m^{-1} $, comparing to the canonical self-resonance ending at $t \sim 150 m^{-1} $.

\begin{figure}
\centering
\subfigure []{\includegraphics[width=5.5cm]{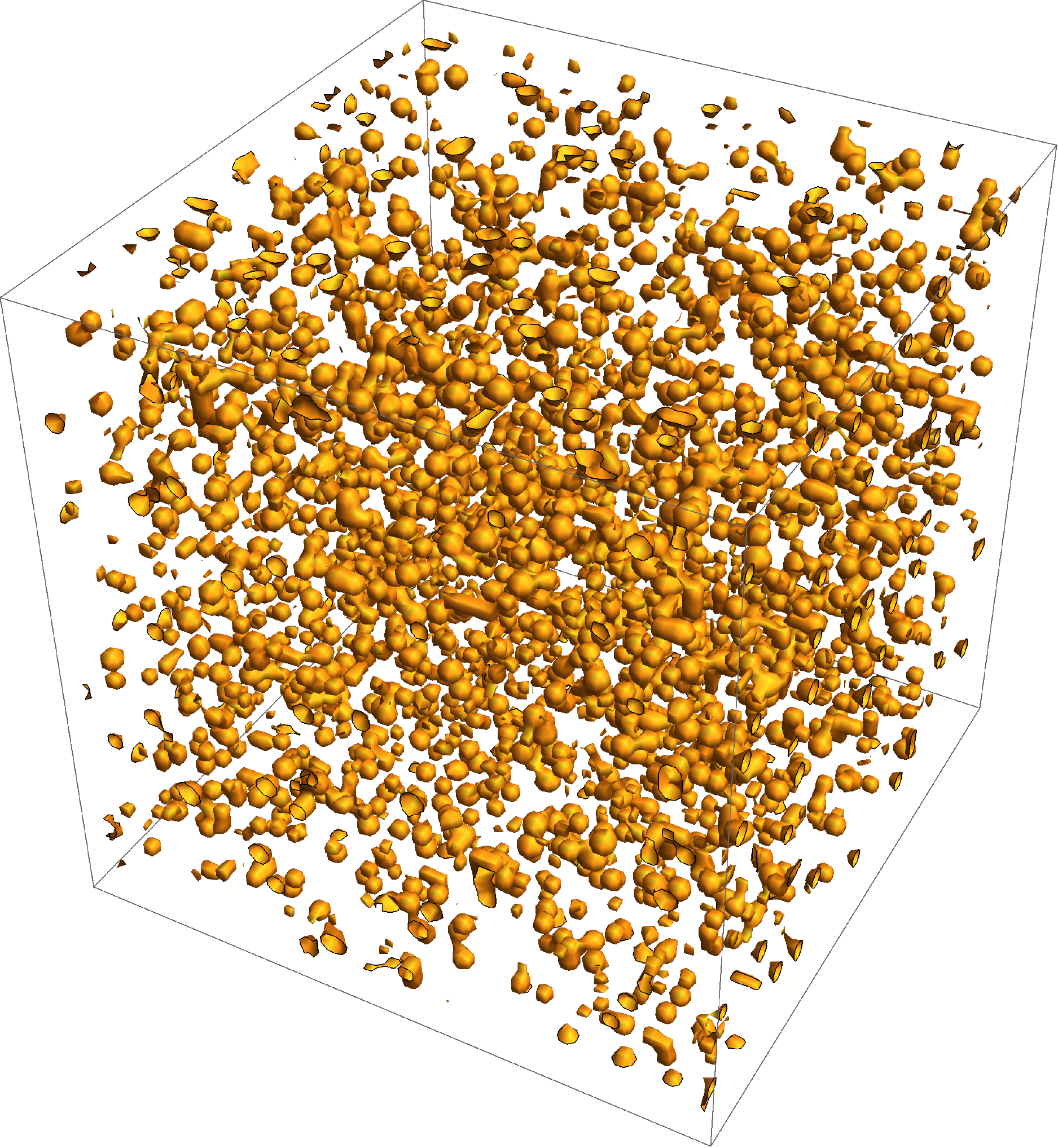}}
\\
\centering
\subfigure[ ]{\includegraphics[width=5.5cm]{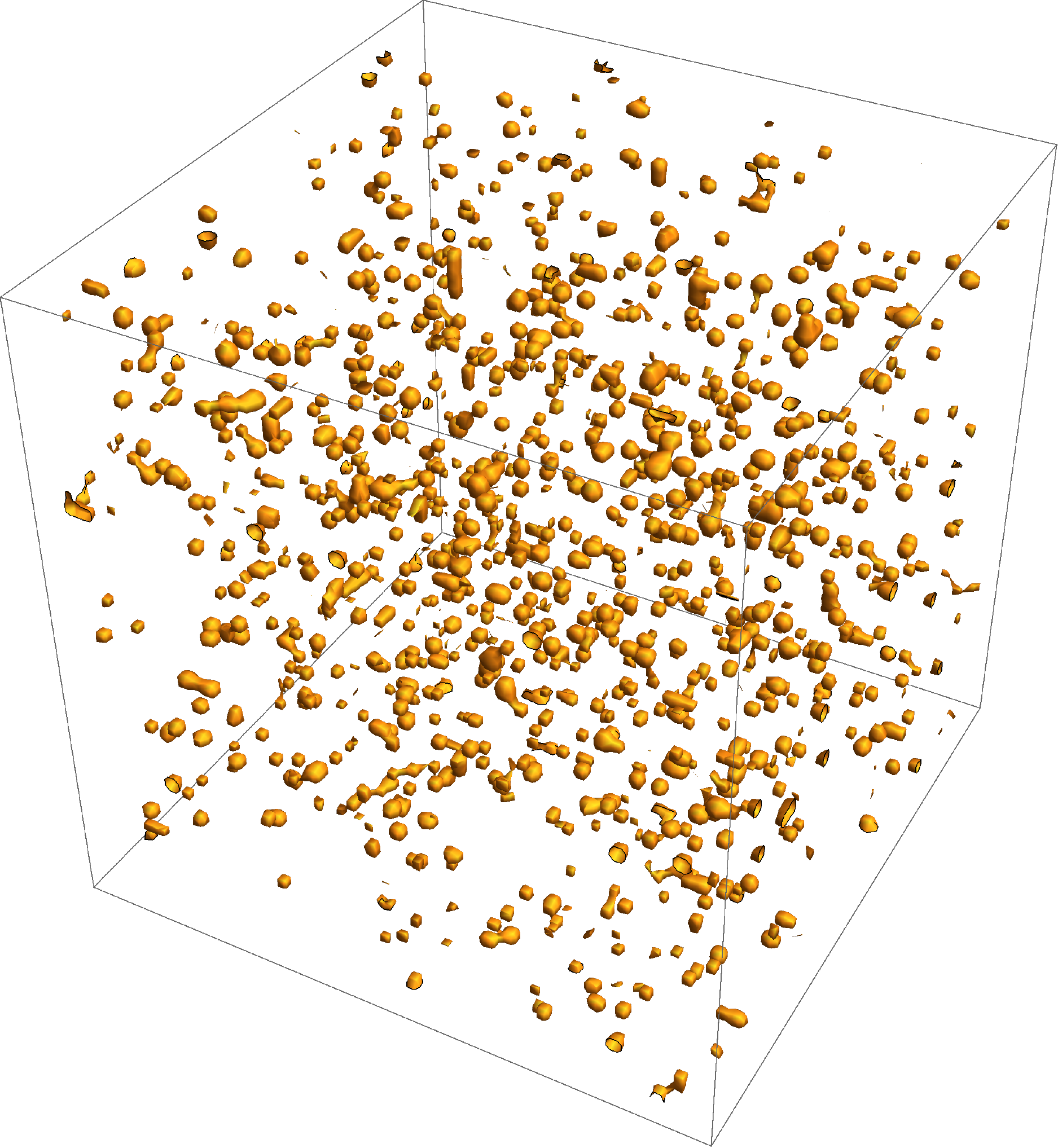}}
\\
\centering
\subfigure[ ]{\includegraphics[width=5.5cm]{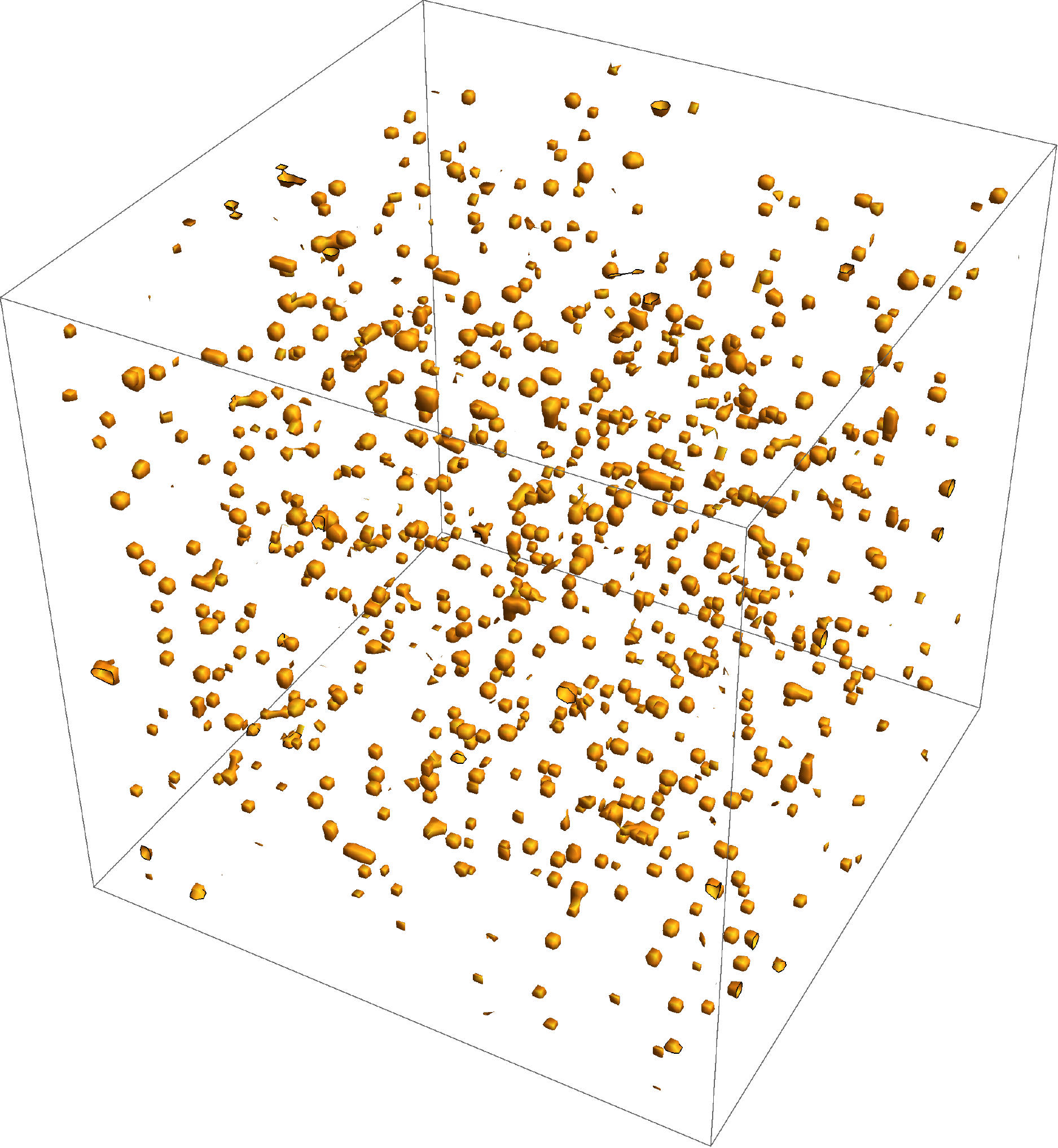}}
\caption{
The snapshot of the energy density at $t = 300 m^{-1} $ for simulations with 
(a) canonical case $\gamma = 1$ (upper panel),   
(b) $\mu = 5\times 10^3 M_{\rm{pl}}$, $\lambda=3\times10^{26}$, $\gamma_{\rm{max}} \approx 2.075$ (middle panel), 
and (c) $\mu = 5\times 10^3 M_{\rm{pl}}$, $\lambda=1.0 \times 10^{27}$, $\gamma_{\rm{max}} \approx 3.898$ (lower panel).
The energy density isosurface is taken at $\rho = 5\langle \rho \rangle$.
}
\label{fig:osc} 
\end{figure}

\begin{figure}
\centering
\subfigure []{\includegraphics[width=5.5cm]{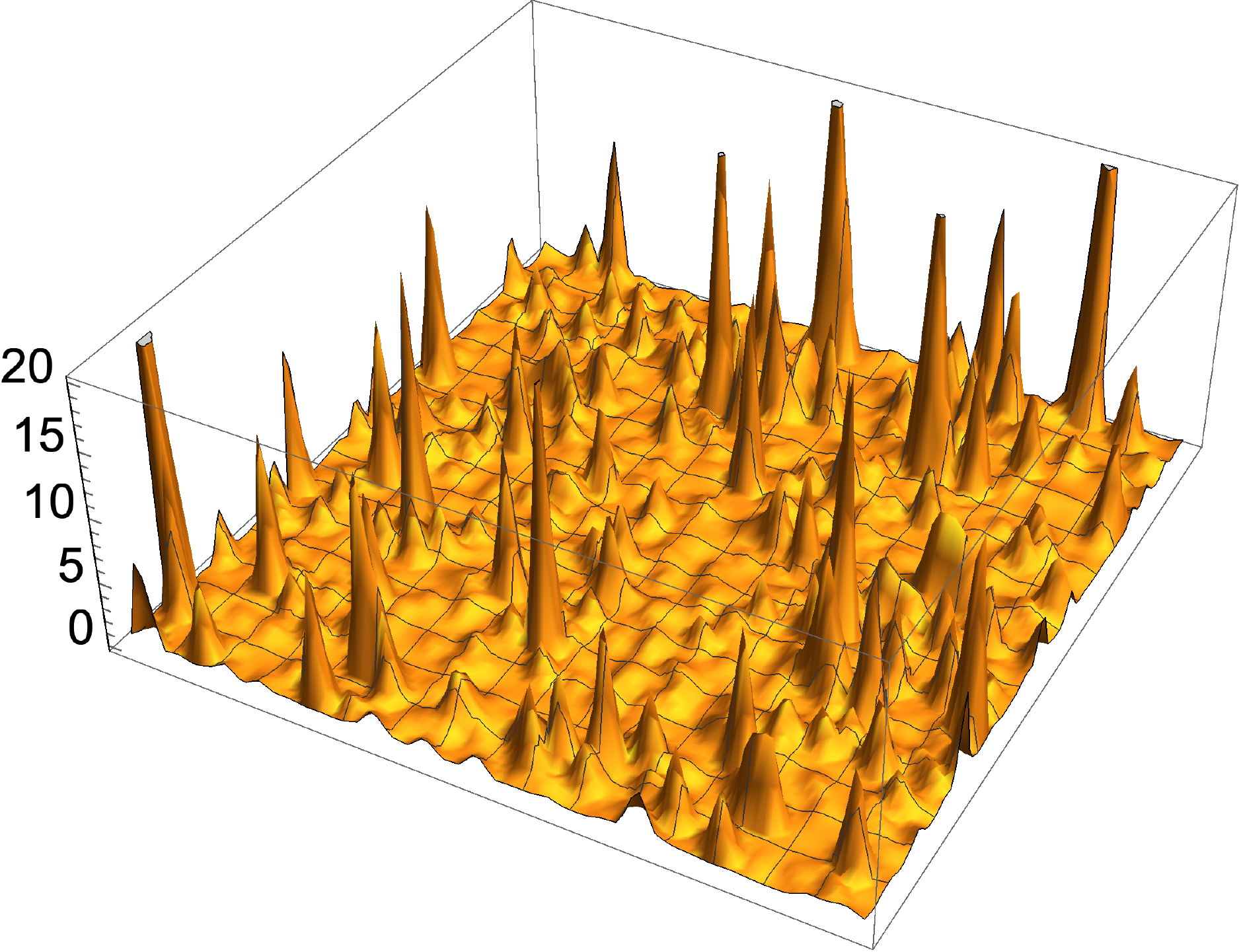}} 
\\
\centering
\subfigure[ ]{\includegraphics[width=5.5cm]{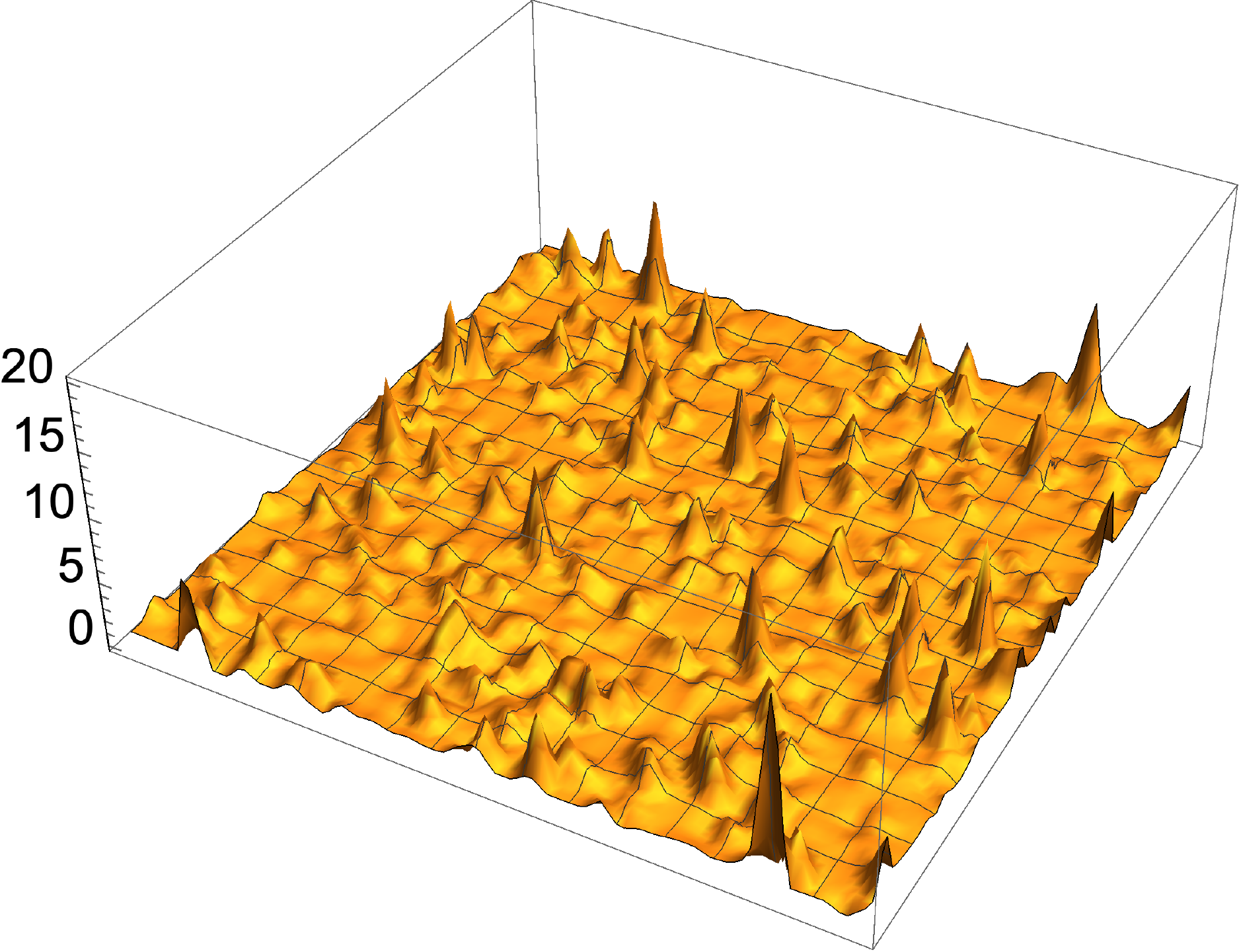}}
\\
\centering
\subfigure[ ]{\includegraphics[width=5.5cm]{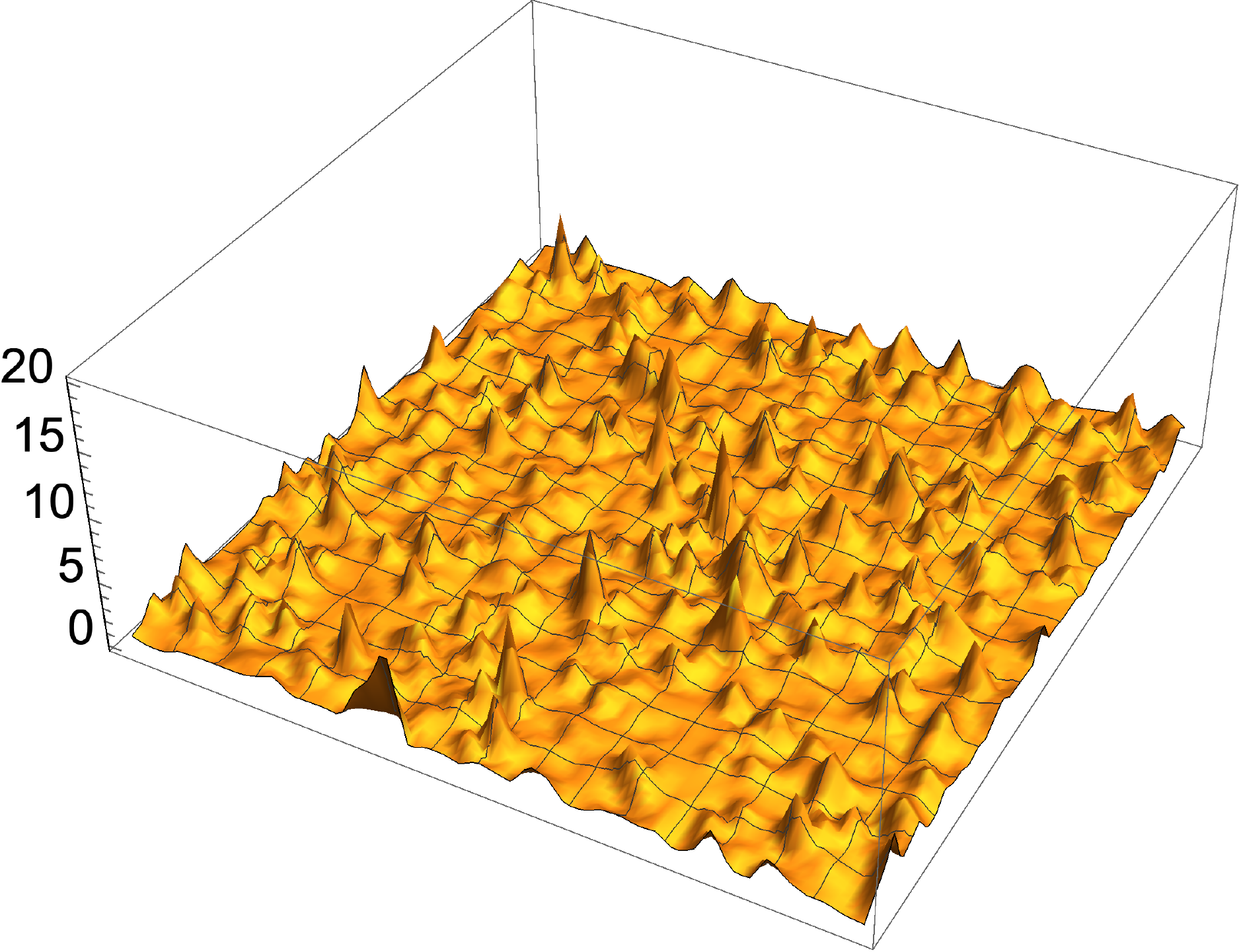}}
\caption{
The snapshot of the energy density $\rho / \langle \rho \rangle$ on a two-dimensional slice at $t = 300 m^{-1} $ for simulations with 
(a) canonical case $\gamma = 1$ (upper panel), 
(b) $\mu = 5\times 10^3 M_{\rm{pl}}$, $\lambda=3\times10^{26}$, $\gamma_{\rm{max}} \approx 2.075$ (middle panel),
and (c) $\mu = 5\times 10^3 M_{\rm{pl}}$, $\lambda=1.0 \times 10^{27}$, $\gamma_{\rm{max}} \approx 3.898$ (lower panel).
} 
\label{fig:2d_osc} 
\end{figure}

The oscillon configurations at $t = 300 m^{-1} $ are shown in Fig.~\ref{fig:osc} and Fig.~\ref{fig:2d_osc}. 
The number of oscillons does not change with time in our simulations, corresponding to an oscillon dominated phase when oscillon has been fully formed. In both figures, the canonical case is shown in the upper panel, the noncanonical case with $\mu = 5\times 10^3 M_{\rm{pl}}$ and $\lambda=3\times10^{26}$ is shown in the middle panel, and the noncanonical case with $\mu = 5\times 10^3 M_{\rm{pl}}$ and $\lambda=1.0 \times 10^{27}$ is shown in the lower panel. In Fig.~\ref{fig:osc}  the energy density isosurface is taken at a value 5 times the average energy density over lattice. The number of the overdensity region and the volume for each one in canonical model are much larger than those in noncanonical cases. Fig.~\ref{fig:2d_osc} is a two-dimensional slice of the energy density rescaled with the average value. The energy density in some region is even up to 40$\langle \rho \rangle$ in the canonical case, but it is suppressed to only 15$\langle \rho \rangle$ in the noncanonical case of $\lambda=3\times10^{26}$ and 8$\langle \rho \rangle$ in the noncanonical case of $\lambda=1.0 \times 10^{27}$.

{\it Discussion.} 
In the canonical case, oscillons can form when the potential satisfies the \textit{open up} condition, i.e., the potential is quadratic at the bottom, and is shallower than quadratic in the field space away from the minimum. For the case in this letter, the contribution of potential terms on the nonlinear evolution of field is reduced by the noncanonical kinetic terms, leading to the suppression of self-resonance and oscillon formation. Our result is derived from numerical simulation using $(3+1)$-dimensional lattice without any approximation of small amplitude oscillons. In fact, Ref. \cite{Amin:2013ika} studies a rather general form of scalar field Lagrangians with noncanonical kinetic terms and gives the condition for the existence of oscillons in the small amplitude limit, and they supposed that the existence of oscillons can be supported solely by the noncanonical kinetic terms, without any need for nonlinear potential terms. However the conclusion for small amplitude oscillons cannot be generalized to large amplitude case.

The choice of values of parameters in the model needs further studied in the further works. As discussed in Ref. \cite{Amin:2011hj}, the formation of oscillon is sensitive to parameter $M$ in potential (\ref{eq:axionV}) but is insensitive to the detailed form of the potential. For the parameters in DBI model, we expect the influence of oscillons is more efficient when the parameters depart further from the canonical case. 

The gravitational-wave background during the formation of oscillon has not been simulated in this work. Due to the presence of the noncanonical effect in the model, we expect the gravitational-wave spectrum should be significantly different from the canonical model as well. We hope to figure out this issue in the near future.

{\it Acknowledgments. }
We would like to thank John T. Giblin and Xue Zhang for helpful discussion. 
We acknowledge the use of HPC Cluster of ITP-CAS. 
Y. S. is supported by grants from NSFC (Grant No. 12005184). 
Q. G. H is supported by grants from NSFC (grant No. 11975019, 11690021, 11991052, 12047503),  the Key Research Program of the Chinese Academy of Sciences (Grant NO. XDPB15), Key Research Program of Frontier Sciences, CAS, Grant NO. ZDBS-LY-7009, CAS Project for Young Scientists in Basic Research YSBR-006




\begin{thebibliography}{99}
\frenchspacing


\bibitem{Bogolyubsky:1976yu}
I.~L.~Bogolyubsky and V.~G.~Makhankov,
Lifetime of Pulsating Solitons in Some Classical Models,
Pisma Zh. Eksp. Teor. Fiz. \textbf{24}, 15-18 (1976)

\bibitem{Gleiser:1993pt}
M.~Gleiser,
Pseudostable bubbles,
Phys. Rev. D \textbf{49}, 2978-2981 (1994)
[arXiv:hep-ph/9308279 [hep-ph]].

\bibitem{Copeland:1995fq}
E.~J.~Copeland, M.~Gleiser and H.~R.~Muller,
Oscillons: Resonant configurations during bubble collapse,
Phys. Rev. D \textbf{52}, 1920-1933 (1995)
[arXiv:hep-ph/9503217 [hep-ph]].

\bibitem{Copeland:2002ku}
E.~J.~Copeland, S.~Pascoli and A.~Rajantie,
Dynamics of tachyonic preheating after hybrid inflation,
Phys. Rev. D \textbf{65}, 103517 (2002)
[arXiv:hep-ph/0202031 [hep-ph]].

\bibitem{Gleiser:2006te}
M.~Gleiser,
Oscillons in scalar field theories: Applications in higher dimensions and inflation,
Int. J. Mod. Phys. D \textbf{16}, 219-229 (2007)
[arXiv:hep-th/0602187 [hep-th]].

\bibitem{Hindmarsh:2006ur}
M.~Hindmarsh and P.~Salmi,
Numerical investigations of oscillons in 2 dimensions,
Phys. Rev. D \textbf{74}, 105005 (2006)
[arXiv:hep-th/0606016 [hep-th]].

\bibitem{Saffin:2006yk}
P.~M.~Saffin and A.~Tranberg,
Oscillons and quasi-breathers in D+1 dimensions,
JHEP \textbf{01}, 030 (2007)
[arXiv:hep-th/0610191 [hep-th]].

\bibitem{Graham:2006vy}
N.~Graham,
An Electroweak oscillon,
Phys. Rev. Lett. \textbf{98}, 101801 (2007)
[erratum: Phys. Rev. Lett. \textbf{98}, 189904 (2007)]
[arXiv:hep-th/0610267 [hep-th]].

\bibitem{Gleiser:2008ty}
M.~Gleiser and D.~Sicilia,
Analytical Characterization of Oscillon Energy and Lifetime,
Phys. Rev. Lett. \textbf{101}, 011602 (2008)
[arXiv:0804.0791 [hep-th]].

\bibitem{Amin:2010jq}
M.~A.~Amin and D.~Shirokoff,
Flat-top oscillons in an expanding universe,
Phys. Rev. D \textbf{81}, 085045 (2010)
[arXiv:1002.3380 [astro-ph.CO]].

\bibitem{Amin:2010dc}
M.~A.~Amin, R.~Easther and H.~Finkel,
Inflaton Fragmentation and Oscillon Formation in Three Dimensions,
JCAP \textbf{12}, 001 (2010)
[arXiv:1009.2505 [astro-ph.CO]].

\bibitem{Amin:2011hj} 
  M.~A.~Amin, R.~Easther, H.~Finkel, R.~Flauger and M.~P.~Hertzberg,
  Oscillons After Inflation,
  Phys.\ Rev.\ Lett.\  {\bf 108}, 241302 (2012)
  [arXiv:1106.3335 [astro-ph.CO]].

\bibitem{Antusch:2017flz} 
  S.~Antusch, F.~Cefala, S.~Krippendorf, F.~Muia, S.~Orani and F.~Quevedo,
  Oscillons from String Moduli,
  JHEP {\bf 1801}, 083 (2018)
  [arXiv:1708.08922 [hep-th]].
  
 \bibitem{Antusch:2019qrr}
S.~Antusch, F.~Cefal\`a and F.~Torrent\'\i{},
Properties of Oscillons in Hilltop Potentials: energies, shapes, and lifetimes,
JCAP \textbf{10}, 002 (2019)
[arXiv:1907.00611 [hep-ph]].

\bibitem{Kou:2019bbc}
X.~X.~Kou, C.~Tian and S.~Y.~Zhou,
Oscillon Preheating in Full General Relativity,
Class. Quant. Grav. \textbf{38}, no.4, 045005 (2021)
[arXiv:1912.09658 [gr-qc]].

\bibitem{Nazari:2020fmk}
Z.~Nazari, M.~Cicoli, K.~Clough and F.~Muia,
Oscillon collapse to black holes,
[arXiv:2010.05933 [gr-qc]].



\bibitem{Gleiser:2010qt}
M.~Gleiser, N.~Graham and N.~Stamatopoulos,
Long-Lived Time-Dependent Remnants During Cosmological Symmetry Breaking: From Inflation to the Electroweak Scale,
Phys. Rev. D \textbf{82}, 043517 (2010)
[arXiv:1004.4658 [astro-ph.CO]].

\bibitem{Gleiser:2011xj}
M.~Gleiser, N.~Graham and N.~Stamatopoulos,
Generation of Coherent Structures After Cosmic Inflation,
Phys. Rev. D \textbf{83}, 096010 (2011)
[arXiv:1103.1911 [hep-th]].

\bibitem{Lozanov:2014zfa}
K.~D.~Lozanov and M.~A.~Amin,
End of inflation, oscillons, and matter-antimatter asymmetry,
Phys. Rev. D \textbf{90}, no.8, 083528 (2014)
[arXiv:1408.1811 [hep-ph]].

\bibitem{Lozanov:2016hid}
K.~D.~Lozanov and M.~A.~Amin,
Equation of State and Duration to Radiation Domination after Inflation,
Phys. Rev. Lett. \textbf{119}, no.6, 061301 (2017)
[arXiv:1608.01213 [astro-ph.CO]].

\bibitem{Lozanov:2017hjm}
K.~D.~Lozanov and M.~A.~Amin,
Self-resonance after inflation: oscillons, transients and radiation domination,
Phys. Rev. D \textbf{97}, no.2, 023533 (2018)
[arXiv:1710.06851 [astro-ph.CO]].

\bibitem{Lozanov:2019ylm} 
  K.~D.~Lozanov and M.~A.~Amin,
  Gravitational perturbations from oscillons and transients after inflation,
  Phys.\ Rev.\ D {\bf 99}, 123504 (2019)
  [arXiv:1902.06736 [astro-ph.CO]].  



\bibitem{Zhou:2013tsa} 
  S.~Y.~Zhou, E.~J.~Copeland, R.~Easther, H.~Finkel, Z.~G.~Mou and P.~M.~Saffin,
  Gravitational Waves from Oscillon Preheating,
  JHEP {\bf 1310}, 026 (2013)
  [arXiv:1304.6094 [astro-ph.CO]].

\bibitem{Antusch:2016con}
S.~Antusch, F.~Cefala and S.~Orani,
Gravitational waves from oscillons after inflation,
Phys. Rev. Lett. \textbf{118}, no.1, 011303 (2017)
[erratum: Phys. Rev. Lett. \textbf{120}, no.21, 219901 (2018)]
[arXiv:1607.01314 [astro-ph.CO]].

\bibitem{Antusch:2017vga}
S.~Antusch, F.~Cefala and S.~Orani,
What can we learn from the stochastic gravitational wave background produced by oscillons?,
JCAP \textbf{03}, 032 (2018)
[arXiv:1712.03231 [astro-ph.CO]].
    
\bibitem{Liu:2017hua} 
  J.~Liu, Z.~K.~Guo, R.~G.~Cai and G.~Shiu,
  Gravitational Waves from Oscillons with Cuspy Potentials,
  Phys.\ Rev.\ Lett.\  {\bf 120}, no. 3, 031301 (2018)
  [arXiv:1707.09841 [astro-ph.CO]].
  
\bibitem{Liu:2018rrt} 
  J.~Liu, Z.~K.~Guo, R.~G.~Cai and G.~Shiu,
  Gravitational wave production after inflation with cuspy potentials,
  Phys.\ Rev.\ D {\bf 99}, no. 10, 103506 (2019)
  [arXiv:1812.09235 [astro-ph.CO]].  
  
\bibitem{Amin:2018xfe} 
  M.~A.~Amin, J.~Braden, E.~J.~Copeland, J.~T.~Giblin, C.~Solorio, Z.~J.~Weiner and S.~Y.~Zhou,
  Gravitational waves from asymmetric oscillon dynamics?,
  Phys.\ Rev.\ D {\bf 98}, 024040 (2018)
  [arXiv:1803.08047 [astro-ph.CO]].
  
\bibitem{Sang:2019ndv} 
  Y.~Sang and Q.~G.~Huang,
  Stochastic Gravitational-Wave Background from Axion-Monodromy Oscillons in String Theory During Preheating,
  Phys.\ Rev.\ D {\bf 100}, no. 6, 063516 (2019)
  [arXiv:1905.00371 [astro-ph.CO]].

\bibitem{Hiramatsu:2020obh}
T.~Hiramatsu, E.~I.~Sfakianakis and M.~Yamaguchi,
Gravitational wave spectra from oscillon formation after inflation,
JHEP \textbf{03}, 021 (2021)
[arXiv:2011.12201 [hep-ph]].


\bibitem{Dvali:1998pa}
G.~R.~Dvali and S.~H.~H.~Tye,
Brane inflation,
Phys. Lett. B \textbf{450}, 72-82 (1999)
[arXiv:hep-ph/9812483 [hep-ph]].

\bibitem{Dvali:2001fw}
G.~R.~Dvali, Q.~Shafi and S.~Solganik,
D-brane inflation,
[arXiv:hep-th/0105203 [hep-th]].

\bibitem{Kachru:2003sx}
S.~Kachru, R.~Kallosh, A.~D.~Linde, J.~M.~Maldacena, L.~P.~McAllister and S.~P.~Trivedi,
Towards inflation in string theory,
JCAP \textbf{10}, 013 (2003)
[arXiv:hep-th/0308055 [hep-th]].

\bibitem{Burgess:2001fx}
C.~P.~Burgess, M.~Majumdar, D.~Nolte, F.~Quevedo, G.~Rajesh and R.~J.~Zhang,
The Inflationary brane anti-brane universe,
JHEP \textbf{07}, 047 (2001)
[arXiv:hep-th/0105204 [hep-th]].

\bibitem{Shiu:2001sy}
G.~Shiu and S.~H.~H.~Tye,
Some aspects of brane inflation,
Phys. Lett. B \textbf{516}, 421-430 (2001)
[arXiv:hep-th/0106274 [hep-th]].



\bibitem{Lachapelle:2008sy}
J.~Lachapelle and R.~H.~Brandenberger,
Preheating with Non-Standard Kinetic Term,
JCAP \textbf{04}, 020 (2009)
[arXiv:0808.0936 [hep-th]].

\bibitem{Matsuda:2008hk}
T.~Matsuda,
Non-standard kinetic term as a natural source of non-Gaussianity,
JHEP \textbf{10}, 089 (2008)
[arXiv:0810.3291 [hep-ph]].

\bibitem{Bouatta:2010bp}
N.~Bouatta, A.~C.~Davis, R.~H.~Ribeiro and D.~Seery,
Preheating in Dirac-Born-Infeld inflation,
JCAP \textbf{09}, 011 (2010)
[arXiv:1005.2425 [astro-ph.CO]].

\bibitem{Karouby:2011xs}
J.~Karouby, B.~Underwood and A.~C.~Vincent,
Preheating with the Brakes On: The Effects of a Speed Limit,
Phys. Rev. D \textbf{84}, 043528 (2011)
[arXiv:1105.3982 [hep-th]].

\bibitem{Child:2013ria}
H.~L.~Child, J.~T.~Giblin, Jr, R.~H.~Ribeiro and D.~Seery,
Preheating with Non-Minimal Kinetic Terms,
Phys. Rev. Lett. \textbf{111}, 051301 (2013)
[arXiv:1305.0561 [astro-ph.CO]].



\bibitem{Amin:2013ika}
M.~A.~Amin,
K-oscillons: Oscillons with noncanonical kinetic terms,
Phys. Rev. D \textbf{87}, no.12, 123505 (2013)
[arXiv:1303.1102 [astro-ph.CO]].



\bibitem{Silverstein:2003hf}
E.~Silverstein and D.~Tong,
Scalar speed limits and cosmology: Acceleration from D-cceleration,
Phys. Rev. D \textbf{70}, 103505 (2004)
[arXiv:hep-th/0310221 [hep-th]].

\bibitem{Alishahiha:2004eh}
M.~Alishahiha, E.~Silverstein and D.~Tong,
DBI in the sky,
Phys. Rev. D \textbf{70}, 123505 (2004)
[arXiv:hep-th/0404084 [hep-th]].



\bibitem{McAllister:2008hb} 
  L.~McAllister, E.~Silverstein and A.~Westphal,
  Gravity Waves and Linear Inflation from Axion Monodromy,
  Phys.\ Rev.\ D {\bf 82}, 046003 (2010)
  [arXiv:0808.0706 [hep-th]].

\bibitem{Silverstein:2008sg} 
  E.~Silverstein and A.~Westphal,
  Monodromy in the CMB: Gravity Waves and String Inflation,
  Phys.\ Rev.\ D {\bf 78}, 106003 (2008)
  [arXiv:0803.3085 [hep-th]].



\bibitem{GABEwebpage}
 http://cosmo.kenyon.edu/gabe.html

\bibitem{Deskins:2013dwa}
J.~T.~Deskins, J.~T.~Giblin and R.~R.~Caldwell,
Gauge Field Preheating at the End of Inflation,
Phys. Rev. D \textbf{88}, no.6, 063530 (2013)
[arXiv:1305.7226 [astro-ph.CO]].

\bibitem{Adshead:2015pva}
P.~Adshead, J.~T.~Giblin, T.~R.~Scully and E.~I.~Sfakianakis,
Gauge-preheating and the end of axion inflation,
JCAP \textbf{12}, 034 (2015)
[arXiv:1502.06506 [astro-ph.CO]].

\bibitem{Adshead:2017xll}
P.~Adshead, J.~T.~Giblin and Z.~J.~Weiner,
Non-Abelian gauge preheating,
Phys. Rev. D \textbf{96}, no.12, 123512 (2017)
[arXiv:1708.02944 [hep-ph]].

\bibitem{Nguyen:2019kbm}
R.~Nguyen, J.~van de Vis, E.~I.~Sfakianakis, J.~T.~Giblin and D.~I.~Kaiser,
Nonlinear Dynamics of Preheating after Multifield Inflation with Nonminimal Couplings,
Phys. Rev. Lett. \textbf{123}, no.17, 171301 (2019)
[arXiv:1905.12562 [hep-ph]].

\bibitem{Giblin:2019nuv}
J.~T.~Giblin and A.~J.~Tishue,
Preheating in Full General Relativity,
Phys. Rev. D \textbf{100}, no.6, 063543 (2019)
[arXiv:1907.10601 [gr-qc]].



\bibitem{Frolov:2008hy}
A.~V.~Frolov,
DEFROST: A New Code for Simulating Preheating after Inflation,
JCAP \textbf{11}, 009 (2008)
[arXiv:0809.4904 [hep-ph]].



 


\end{thebibliography}
\end{document}